\documentclass{optica-article}

\journal{opticajournal} 

\articletype{Research Article}
\usepackage{todonotes}
\usepackage{siunitx}

\usepackage{amsmath}

\begin{document}

\title{qtOCT: quantitative transmission optical coherence tomography}

\author{Wojciech Krauze\authormark{*}, Martyna Mazur, and Arkadiusz Kuś}

\address{Warsaw University of Technology, Institute of Micromechanics and Photonics, Boboli 8 street, 02-525, Warsaw Poland}

\email{\authormark{*}wojciech.krauze@pw.edu.pl} 


\begin{abstract*} 
Transmission optical coherence tomography (OCT) enables analysis of biological specimens \textit{in vitro} through detection of forward scattered light. Up to now, transmission OCT was considered as a technique that cannot directly retrieve quantitative phase and is thus a qualitative method. In this paper, we present qtOCT, a novel quantitative transmission optical coherence tomography method. Unlike existing approaches, qtOCT allows for a direct, easy, fast and rigorous retrieval of 2D integrated phase information from transmission full-field swept-source OCT measurements. Our method is based on coherence gating and allows user-defined temporal measurement range selection, making it potentially suitable for analyzing multiple-scattering samples. We demonstrate high consistency between qtOCT and digital holographic microscopy phase images. This approach enhances transmission OCT capabilities, positioning it as a viable alternative to quantitative phase imaging techniques.
\end{abstract*}

\section{Introduction}

Optical coherence tomography (OCT) is an imaging technique, most commonly realized in reflection configuration where the sample is illuminated in the \textit{epi} mode \cite{drexler2008optical}. It allows retrieving qualitative information about refractive index gradients with relatively high axial resolution \textit{in vivo}, however with refraction-induced distortions. In recent years, transmission OCT (tOCT) systems have gained attention of the scientific community \cite{trull2015transmission}. One reason behind this interest is that in most biological samples forward scattered light is much stronger than the backward scattered light due to their high anisotropy parameter \cite{wilson1990optical}. Additionally, lack of \textit{epi} mode measurement is not an issue when biological \textit{in vitro} measurements are carried out. Unfortunately, despite the fact that the theoretical optical transfer function in tOCT systems is in the low frequency region (covering the DC term) \cite{zhou2021unified}, up to now there is no direct method that can retrieve the phase from such OCT measurements. Indirect methods include localization of the ballistic peak position in point scanning OCT \cite{van2020deep, wang2010high}, which requires very precise control of the optical path difference (OPD) between sample and reference arms of an OCT system and can be applied only where the ballistic peak is clearly visible and thus are not universal and practical. Another indirect approach is based on holoscopy-like systems \cite{zvyagin2005image, potcoava2008optical} which do not take advantage of the OCT imaging properties, like high sensitivity due to cross-correlation of the sample and reference beams, and instead process captured data in a classical interferometry approach. Anna et al. \cite{anna2011transmission} presented a method of retrieving phase images from tOCT images, however it was implemented erroneously, which is confirmed by wrong phase values in the reconstructed images.

In this paper we present the first to our knowledge direct method to retrieve complex amplitudes from measurements captured with full-field swept-source OCT in transmission (t-FF-SS-OCT). Our method, called qtOCT (quantitative transmission OCT), is a novel signal processing method strictly based on OCT theory and as such it does not require any modifications to standard t-FF-SS-OCT systems. We identify 3 main advantages of qtOCT. Firstly, this is a first method where 2D transverse quantitative phase information can be directly retrieved from tOCT data. In that sense, the resulting information can also be referred to as quantitative, integrated \textit{en face} image. Secondly, since our solution is based on coherence gating phenomenon, it allows straightforward selection of user-defined temporal measurement range, which can be used to discard multiple scattered photons and reduce the problem to a weakly scattering one. What is also important, qtOCT is relatively easy to implement: for those who already have a digital holographic microscopy (DHM) system \cite{kemper2008digital}, it is enough to change the light source to a swept-source laser, and for those who have the t-FF-SS-OCT setup it is enough to change the signal processing algorithm.

The paper is structured as follows. In Section \ref{sec:method} we present the details of our method and explain important differences in interpretation of tOCT measurements in comparison to traditional reflection configuration, which have profound consequences. In Section \ref{sec:results and discussion} we present and discuss the measurement results that prove the quantitative nature of qtOCT and its temporal gating property, and finally in Section \ref{sec:conclusions} we give the conclusions. 

\section{Method}
\label{sec:method}
\subsection{Optical setup}
\label{optical_setup}
\begin{figure}[ht!]
    \centering
    \includegraphics[width=.7\textwidth]{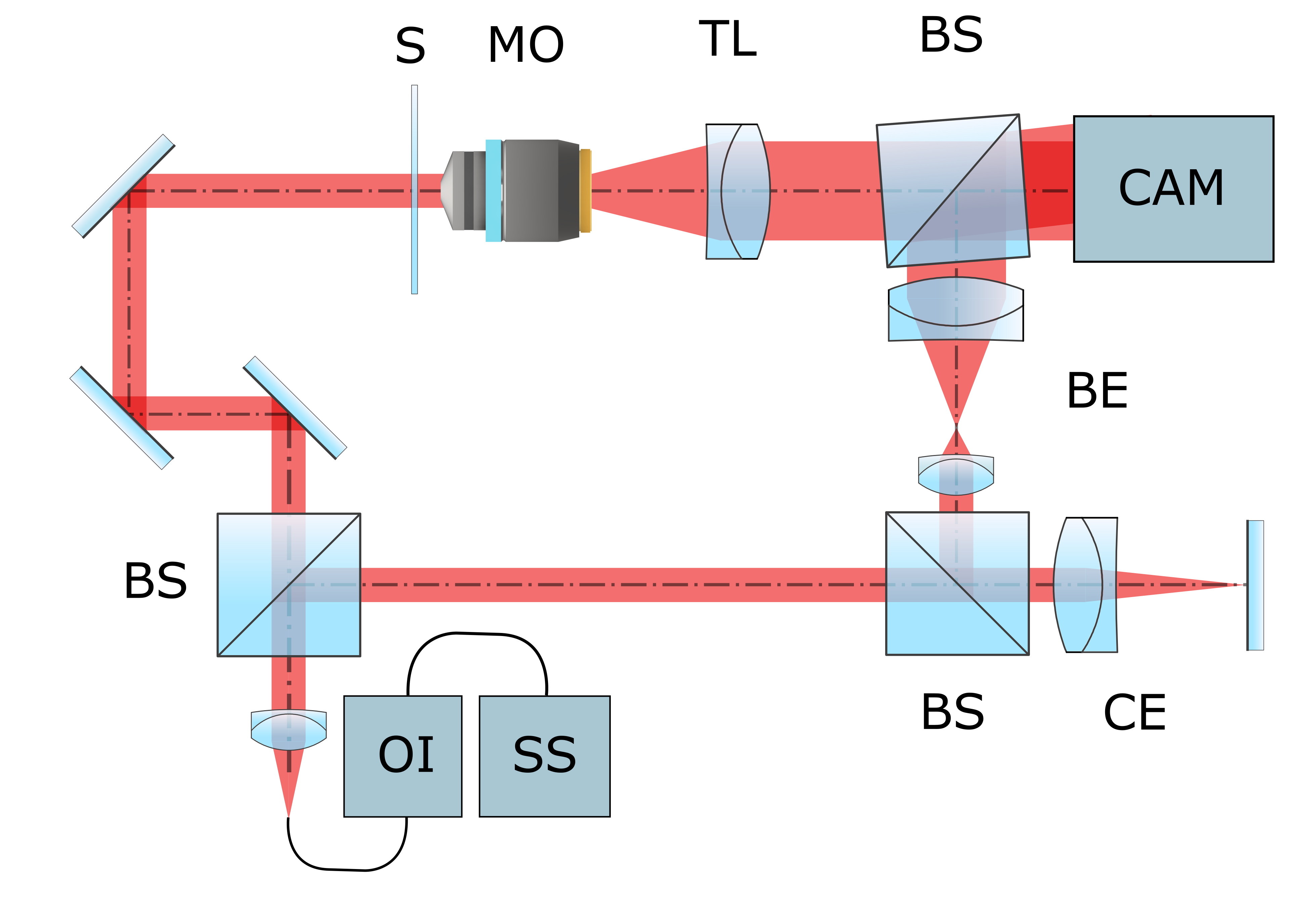}
    \caption{Optical system of t-FF-SS-OCT. SS $-$ swept-source; OI $-$ optical isolator; BS $-$ beam-splitters; CE $-$ cat-eye retroreflector; BE $-$ beam expander; S $-$ sample plane; MO $-$ microscope objective; TL $-$ tube lens; CAM $-$ CMOS camera.}
    \label{fig:optical-system}
\end{figure}

The measurement system used in this work is an off-axis DHM based on Mach-Zehnder interferometer (MZI) setup (Fig.~\ref{fig:optical-system}). The light source is a swept-source (BS-840-1-HP, Superlum) with the full wavelength range of the source $\Delta\lambda$ of \SI{75}{\nano \meter} starting from \SI{803}{\nano \meter} to \SI{878}{\nano \meter} at \SI{20}{\milli \watt} power. The linewidth of the source is \SI{0.06}{\nano \meter}, which results in a coherence length of \SI{10.7}{\milli \meter} at $\lambda =\SI{803}{\nano \meter}$. We use a mirror system in the object beam and a cat-eye retro-reflector (CE in Fig.~\ref{fig:optical-system}) in the reference beam, to have full control over the OPD in the system. The sample is illuminated by a beam diameter of \SI{220}{\micro \meter} (at $1/e^{2}$) and imaged by a $40x$ NA $1.3$ microscope objective (Zeiss) and a tube lens (effective focal length EFL=\SI{300}{\micro \meter}), with magnification of $\text{M}=-71.775$ (measured separately). The beam expander (BE) is used to match the diameter of the reference beam to the object beam at the detector plane. The detector used in this system is a monochromatic CMOS camera (acA2040-180-km, Basler) with \SI{5.5}{\micro \meter} pixels arranged in a $2048x2048$ array, offering a maximum framerate of $180$ fps, which puts the limit on the measurement speed as the source offers a \SI[quotient-mode=fraction, fraction-function=\dfrac]{10000}{\nano \meter / \second} scan speed, which translates to $133.3$ full sweeps per seconds. 

The swept-source provides nearly flat spectral power distribution, thus axial temporal resolution can be defined as $\delta_{t} = \lambda_{c}^{2}/(\Delta \lambda c)=\SI{31}{\femto \second}$. Here, $\lambda_{c}$ and $c$ indicate center wavelength and speed of light, respectively. The lateral resolution is \SI{0.32}{\micro \meter}.

\subsection{Signal processing}

\begin{figure}
    \centering
    \includegraphics[width=.99\textwidth]{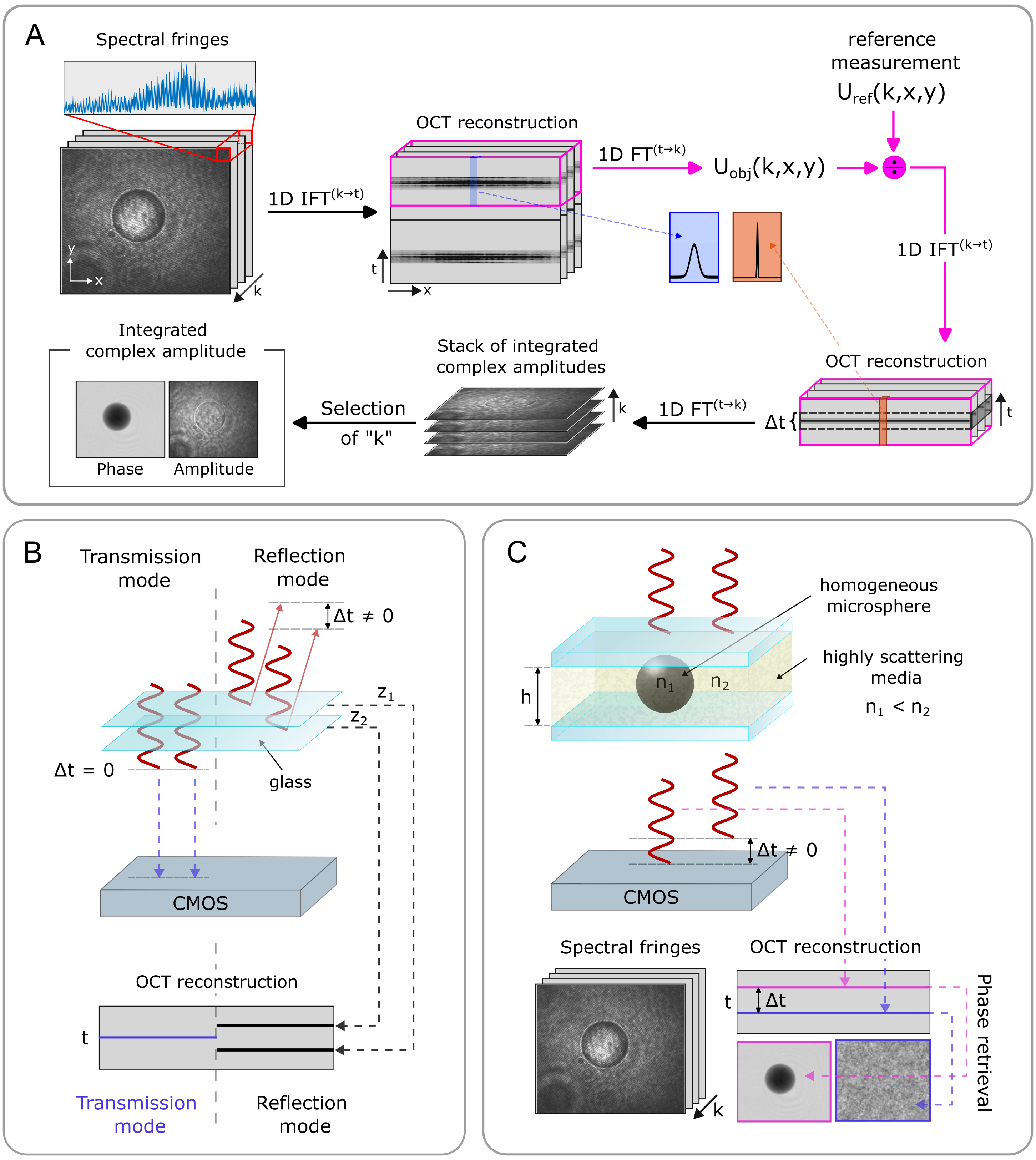}
    \caption{Visualization of the qtOCT principle: A) processing pipeline; B) differences in OCT signal formation in transmission (here qtOCT method) and reflection mode using flat plate with two surfaces \(z_{1}\) and \(z_{2}\) as an investigated sample; C) presentation of the temporal gating in qtOCT. k $-$ wavenumber; FT, IFT $-$ Fourier transform and inverse Fourier transform, respectively; \(\text{U}_{\text{obj}}\) $-$ complex amplitudes before reference correction; \(\Delta t\) $-$ time delay between signals; \(n_{1}\), \(n_{2}\) $-$ refractive indices; h $-$ thickness of the sample. }
    \label{fig:graph}
\end{figure}

Two t-FF-SS-OCT datasets are captured for $N_k$ wavenumbers: one object and one reference measurement. After that, qtOCT processing is applied. Graphical visualization of the processing pipeline is shown in Fig. \ref{fig:graph}A. The tOCT signal captured for each wavenumber $k$ is a real-valued signal in the form of spectral interference fringes and can be described by \cite{wojtkowski2009obrazowanie}:
\begin{equation}
\begin{split}
    I(k) &= \kappa_{\text{ref}}I_0(k)+\kappa_{\text{ob}}I_0(k)\text{FT}^{t'\rightarrow k}\{ \Phi(c t') \circledast \Phi^*(c t') \}+\\
    &+\sqrt{\kappa_{\text{ob}}\kappa_{\text{ref}}}I_0(k)\text{FT}^{t\rightarrow k}\{ \Phi (c t) \}+
    \sqrt{\kappa_{\text{ob}}\kappa_{\text{ref}}}I_0(k)\text{FT}^{t\rightarrow k}\{ \Phi (c t) \}^*
\end{split}
\label{eq:spectral fringes}
\end{equation}
where \(I_0(k)\) is a light source spectrum, \( \kappa_{\text{ref}}\), \( \kappa_{\text{ob}}\) represent spectral scaling factors for the reference and object arms, respectively, \(c\) is a speed of light, \(t'\) characterize transmission time of the sample signal and \(t\) is the time delay between reference and sample signals. The operators \(FT\), \(\circledast\), \(^*\) represent Fourier transform, convolution and complex conjugate, respectively.

In order to retrieve information about investigated sample's integrated phase \(\Phi(c t)\), one-dimensional (1D) Fourier transform along \(k\) is performed. The result is a standard OCT reconstruction:
\begin{equation}
\begin{split}
    \text{IFT}^{k\rightarrow t}\{ I(k)\}&=\kappa_{\text{ref}}I_0(c t)+
    \kappa_{\text{ob}}I_0(c t)\text{ACF}\{ \Phi (c t')\}+\\
    &+\sqrt{\kappa_{\text{ob}}\kappa_{\text{ref}}}I_0(c t) \circledast \Phi(c t)+
    \sqrt{\kappa_{\text{ob}}\kappa_{\text{ref}}}I_0(c t) \circledast \Phi(-c t)
\end{split}
\label{eq:oct-recon}
\end{equation}
where IFT is an inverse Fourier transform, \(I_0(c t)\) is a coherence function of the light source and ACF denotes autocorrelation. 

The first 2 components in Eq.~\ref{eq:oct-recon} are DC and autocorrelation terms, and the information about \(\Phi(c t)\) is encoded in the last two complex conjugate cross-correlation components. To select one of them, a window function (CC window) is applied. After windowing, we receive \(\Phi(c t)\) convolved with \(I_0(c t)\):

\begin{equation}
    \text{CC}_{\text{obj}}=\sqrt{\kappa_{\text{ob}}\kappa_{\text{ref}}}I_0(c t) \circledast \Phi(c t)
\end{equation}
The convolution operation leads to broadening and distortion of \(\Phi(c t)\) signal (see blue box in Fig.~\ref{fig:graph}A) which decreases achievable resolution and accuracy. To decouple the OCT signal from the light source properties, 1D FT of \(\text{CC}_{\text{obj}}\) is calculated:
\begin{equation}
    \text{U}_{\text{obj}}=\text{FT}^{t\rightarrow k}\{ CC_{\text{obj}}\}=\text{FT}^{t\rightarrow k}\{ \sqrt{\kappa_{\text{ob}}\kappa_{\text{ref}}}I_0(c t)\} \cdot \text{FT}^{t\rightarrow k}\{ \Phi(c  t)\}
\end{equation}
and the result \(\text{U}_{\text{obj}}\) is divided by the reference measurement \(\text{U}_{\text{ref}}\), which underwent the same signal processing as the object measurement:
\begin{equation}
    \text{U}_{\text{ref}} = \text{FT}^{t\rightarrow k}\{ CC_{\text{ref}}\}=\text{FT}^{t\rightarrow k}\{ \sqrt{\kappa_{\text{ob}}\kappa_{\text{ref}}}I_0(c t)\}
\end{equation}

\begin{equation}
    \text{FT}^{t\rightarrow k}\{ \Phi(c t)\}=\frac{\text U_{\text{obj}}}{\text U_{\text{ref}}}=\frac{\text{FT}^{t\rightarrow k}\{ CC_{\text{obj}}\}}{\text{FT}^{t\rightarrow k}\{ CC_{\text{ref}}\}}
\end{equation}

Then, we calculate 1D IFT to return to the OCT reconstruction space. Due to performed source-correction, the broadening effect is suppressed and the resolution is significantly increased (see blue and orange boxes in Fig.~\ref{fig:graph}A). At this point, the CC window is corrected to keep only the signals for which delays relative to the reference signal are within selected period of time \(\Delta t\), enabling temporal gating property. After that, 1D FT is performed, and finally integrated complex amplitudes for different \(k\) are obtained:

\begin{equation}
    E^{(r)}(k) = \text{exp}(i(kt))\cdot\text{FT}^{t\rightarrow k}\{ \Phi(c t)\}
\end{equation}

It is now possible to select $E^{(r)}$ for a specific $k$ and retrieve integrated phase of a sample with standard approach by using two-argument arctangent function and a 2D phase unwrapping algorithm \cite{herraez_fast_2002}. 

To give more insight into the operating principle of qtOCT and to show critical differences between transmission and reflection OCT modalities, two examples representing different investigated samples are shown in Fig.~\ref{fig:graph}. As a first example, a flat plate with two surfaces located at different depths \(z_{1}\) and \(z_{2}\) was chosen (see Fig.~\ref{fig:graph}B). Starting the consideration with the \textbf{OCT signal formation in reflection mode}, the incident light travelling trough the sample is backreflected from \(z_{1}\) and \(z_{2}\) surfaces. Due to the fact that \(z_{1} < z_{2}\), the light from \(z_{2}\) surface will reach the detector with a delay of \(\Delta t\) in relation to light coming from \(z_{1}\). The obtained \(\Delta t\) is here determined by the relative location of sample surfaces. This means that in the OCT reconstruction we will see two distinct layers directly corresponding to the depth structure of the flat plate. \textbf{The situation is different in transmission OCT}. The presence of two layers \(z_{1}\) and \(z_{2}\) does not generate time delays (\(\Delta t = 0\)) in transmitted light. The captured signal represents the cumulative information from two surfaces rather than separate information from each of them, resulting in a single layer in the OCT reconstruction. Thus, in transmission OCT, and qtOCT specifically, in such case there is no sectioning property. On the other hand, this also means that qtOCT gives a unique opportunity to provide quantitative \textbf{integrated} phase, similarly to techniques like DHM.

The second example depicted in Fig.~\ref{fig:graph}C further explains the concept of the OCT signal generation in transmission mode. A homogeneous microsphere immersed in a highly scattering medium is now an investigated sample. The object and medium have refractive indices \(n_{1}\) and \(n_{2}\), respectively, and the thickness of the sample is $h$. Let us consider two different light paths: first one through the middle of the homogeneous microsphere (optical path length \(\text{OPL}_{1} = n_{1} \cdot h\)) and the second one through the medium only (\(\text{OPL}_{2} = n_{2} \cdot h\)). Given that \(n_{1} < n_{2}\), \(\text{OPL}_{1} < \text{OPL}_{2}\), and thus the light that travels the first path will reach the detector earlier than light travelling the second path, resulting in temporal delay \(\Delta t\). It should be noted that \(\Delta t\) has to be higher than axial temporal resolution of the qtOCT to receive separation of the two signals in the reconstruction, as presented in Fig.~\ref{fig:graph}C (magenta and blue lines in the OCT reconstruction). Meeting this condition allows to retrieve integrated phase image separately for each signal. This opens up possibilities to perform qtOCT measurements in highly scattering conditions and retrieve integrated phase with discarded multiply scattered photons. Technically, during qtOCT processing this is realized by controlling the value of $\Delta t$ region that is selected to be inverse Fourier-transformed (Fig. \ref{fig:graph}A). A specific $\Delta t$ region results in reconstructed integrated phase that takes into account information carried by all the photons that reach the detector within $\Delta t$. The smallest $\Delta t$ that can be selected is 1 reconstruction plane and due to the fact that 1D FT of 1 sample is the sample itself, in that specific case no IFT is required.

\section{Results and discussion}
\label{sec:results and discussion}

\begin{figure}
    \centering
    \includegraphics[width=\textwidth]{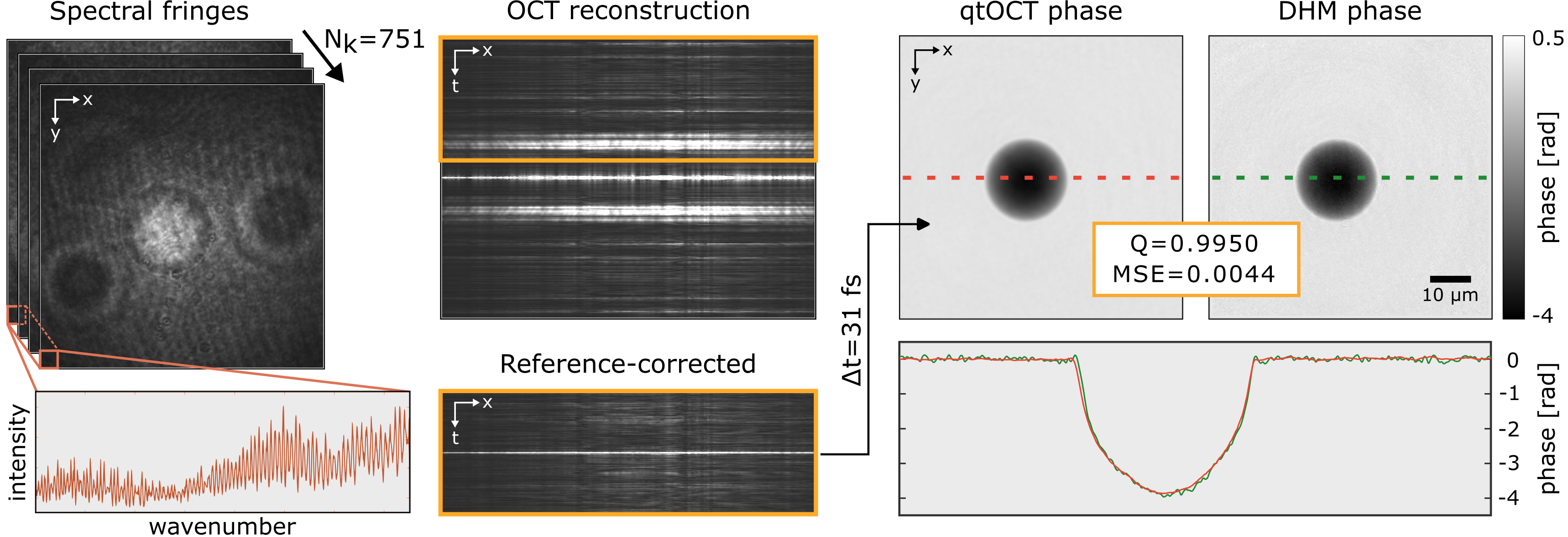}
    \caption{Experiment 1: quantitative comparison of an integrated phase retrieved with qtOCT and DHM. The analyzed spectral fringes are t-FF-SS-OCT holograms converted numerically from off-axis to on-axis configuration (see Supplement 1 for description of the conversion procedure). $\Delta t$ window was used to retrieve the integrated phase with qtOCT. Q $-$ universal image quality index; MSE $-$ mean squared error.}
    \label{fig:exp1}
\end{figure}

\begin{figure}
    \centering
    \includegraphics[width=.8\textwidth]{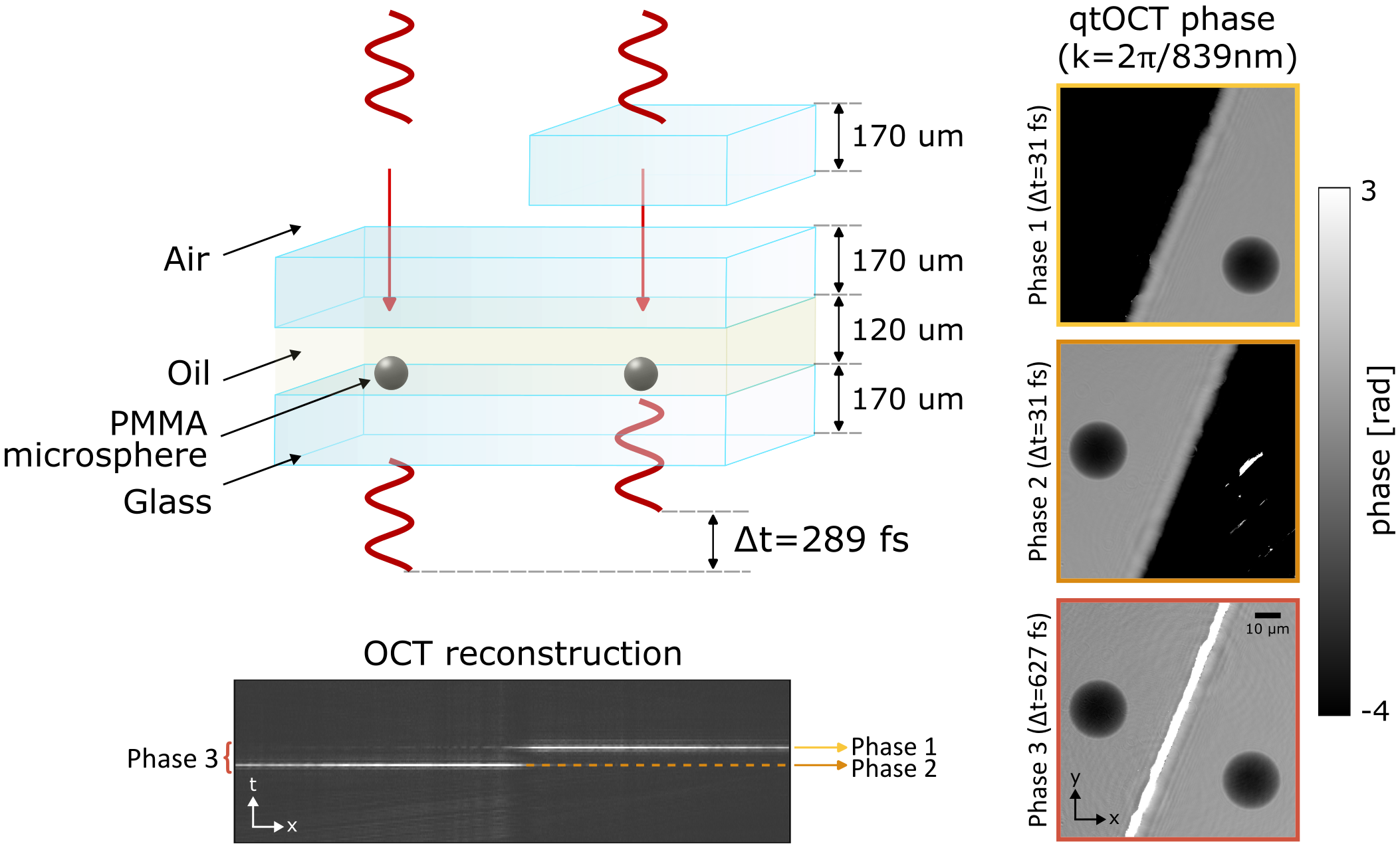}
    \caption{Experiment 2: examination of the temporal gating property of qtOCT. The top glass coverslip covering only one PMMA microsphere was placed in a conjugate plane to the sample plane. The received integrated phases (Phase 1, Phase 2 and Phase 3) have unified background phase values.}
    \label{fig:exp2}
\end{figure}

In this section, the results of two experiments emphasizing different properties of qtOCT approach are presented. Experiment 1 (Fig.~\ref{fig:exp1}) demonstrates the capability of qtOCT to retrieve quantitative integrated phase, while Experiment 2 (Fig.~\ref{fig:exp2}) presents temporal gating property of the proposed approach. In both cases, the t-FF-SS-OCT data was recorded for $751$ wavenumbers using t-FF-SS-OCT system (see Section \ref{optical_setup}). Sample size in OCT reconstructions along t-axis equals theoretical temporal resolution, \SI{31}{\femto \second}.

Starting with the Experiment 1 (Fig.~\ref{fig:exp1}), the investigated specimen was poly(methyl methacrylate) (PMMA) microsphere (microParticles GmbH, $\text{RI}_{839\text{nm}} = 1.4841$) with a diameter of \SI{23.5}{\micro \meter} immersed in Zeiss Immersol $518$F oil ($\text{RI}_{839\text{nm}} = 1.5067$). The tOCT signal obtained for this sample was recorded in off-axis configuration to allow direct comparison of phase measurements with DHM. However, to prove that spatial carrier frequency is not required for qtOCT phase retrieval, off-axis holograms were transformed to on-axis ones, according to the procedure shown in Fig. S1 in Supplementary Materials. The obtained spectral fringes were then processed according to the qtOCT procedure (Fig.~\ref{fig:graph}A). As expected, reference correction increased the resolution of the OCT reconstruction. The strong signal present in reference-corrected OCT reconstruction corresponds to ballistic and weakly scattered photons that reached the detector first. In order to retrieve integrated phase representing the measured microsphere, the captured OCT signal was temporally gated by the smallest possible window $\Delta t = $ \SI{31}{\femto \second}, which is equivalent to a single sample size along $t$-axis. To assess the correctness of phase retrieval, the obtained qtOCT phase was compared to the phase calculated through Fourier-transform algorithm used in DHM. The comparison was performed using two measures: universal image quality index (Q) \cite{wang_universal_2002} and mean squared error (MSE). The resulting Q value of $0.9950$ and MSE equaled to $0.0044$ reveal nearly perfect similarity between qtOCT and DHM phase, which is further confirmed by 1D cross-sections.  

In Experiment 2 (Fig.~\ref{fig:exp2}), the investigated sample was reconfigured. Two PMMA microspheres were located between two glass coverslips, each with a thickness of \SI{170}{\micro \meter}, separated by \SI{120}{\micro \meter} thick spacer. In a plane conjugate to the sample plane, another \SI{170}{\micro \meter} thick coverslip ($\text{RI}_{839\text{nm}} = 1.5100$)  was placed in the field of view in such a way that it covers only one microsphere (for simplicity, the conjugate plane was not depicted in Fig.~\ref{fig:exp2}). The presence of an additional piece of glass resulted in a temporal delay of \SI{289}{\femto \second} between light beams that propagated through the 2 microspheres. Given the available axial temporal resolution of the t-FF-SS-OCT system (\SI{31}{\femto \second}), it is now possible to temporally separate the 2 signals $-$ this is confirmed by spatially separated signals in the OCT reconstruction in Fig.~\ref{fig:exp2}. In order to present temporal gating property of qtOCT, the OCT signal was gated using different time windows \(\Delta t\). For each case, the 2D integrated phases were calculated. 
Phase 1 and Phase 2, corresponding to the parts of the sample with and without additional glass, respectively, were obtained by applying a $\Delta t = $ \SI{31}{\femto \second} window to a specific region of the OCT reconstruction (see Fig.~\ref{fig:exp2}). The obtained phases confirm the correctness of the separation of both signals. In order to retrieve integrated phase information about both microspheres, a wider temporal window is needed that will cover both signals. For this case, we used a time window of $\Delta t= $ \SI{627}{\femto \second}. From the integrated complex amplitudes received, the one for $k = {2\pi}/839\text{nm}$ was selected, and the integrated Phase 3 was calculated, which contains integrated phase values of two separated PMMA microstructures. It should be noted, that due to phase discontinuity at the coverslip edge, the two regions in Phase 3 can be treated independently and their background phase values were unified in order to increase image contrast.

The results derived from both presented experiments confirm two essential features of qtOCT: its quantitative character and ability to gate time-separated t-FF-SS-OCT signals.
Temporal gating property of qtOCT, due to transmission configuration, allows for different applications compared to traditional reflection OCT. It does not allow separating object layers. Instead, it can be used to retrieve integrated quantitative phase of a sample surrounded by a highly-scattering medium by filtering out multiply scattered photons which reach the detector later than weakly scattered ones. The efficiency of this approach depends on the width of the time gate, which in turn depends on the wavelength range of the light source. Here a \SI{75}{\nano \meter}-range swept-source was used, which allows for a minimal time gate of \SI{31}{\femto \second}. This time gate is equivalent to approx. \SI{9}{\micro \meter} OPD between 2 signals in air. 

\section{Conclusions}\label{sec:conclusions}
In this paper a novel qtOCT approach for processing t-FF-SS-OCT data has been presented. The qtOCT is a quantitative method which, as opposed to the already developed tOCT approaches, is capable of recovering 2D integrated phase information in a direct and easy way. We proved high consistency between the phases obtained through our method and DHM. Additionally, we showed that qtOCT can select and analyze time-separated t-FF-SS-OCT signals (in contrast to depth-dependent signals in reflection OCT). This feature makes qtOCT more powerful than other quantitative phase imaging techniques, like DHM and gives the possibility to reduce multiply scattering samples to weakly scattering ones. With ongoing progress of swept-source lasers technology, this approach may allow very precise separation of the incoming photons. Finally, unlike holoscopy or holography-based methods, qtOCT takes advantage of cross-correlation in the signal processing (which enhances the object wave by its interaction with the strong reference wave), which may potentially allow measuring thick or partially absorbing samples that attenuate the object beam.

\section{Backmatter}
\begin{backmatter}
\bmsection{Funding}
Research was funded by POB Photonics of Warsaw University of Technology within the Excellence Initiative: Research University (IDUB) programme and REVEAL project (101016726) from European Union's Horizon 2020 programme.

\bmsection{Author contributions}
\textbf{WK:} Conceptualization, Methodology, Formal analysis, Writing - original draft, Supervision; \textbf{MM:} Methodology, Software, Formal analysis, Writing - original draft, Visualization; \textbf{AK:} Investigation - performing the experiments, Data collection, Writing - description and illustration of the optical system;

\bmsection{Disclosures}
The authors declare no conflicts of interest.

\bmsection{Data availability} Data underlying the results presented in this paper are available in Zenodo repository (https://doi.org/10.5281/zenodo.11124973).

\bmsection{Supplemental document}
See Supplement 1 for supporting content. 

\end{backmatter}

\bibliography{manuscript}

\section*{SUPPLEMENTAL MATERIAL: Generation of on-axis holograms from off-axis holograms} 
\begin{figure}[htbp]
\centering
\includegraphics[width=.85\linewidth]{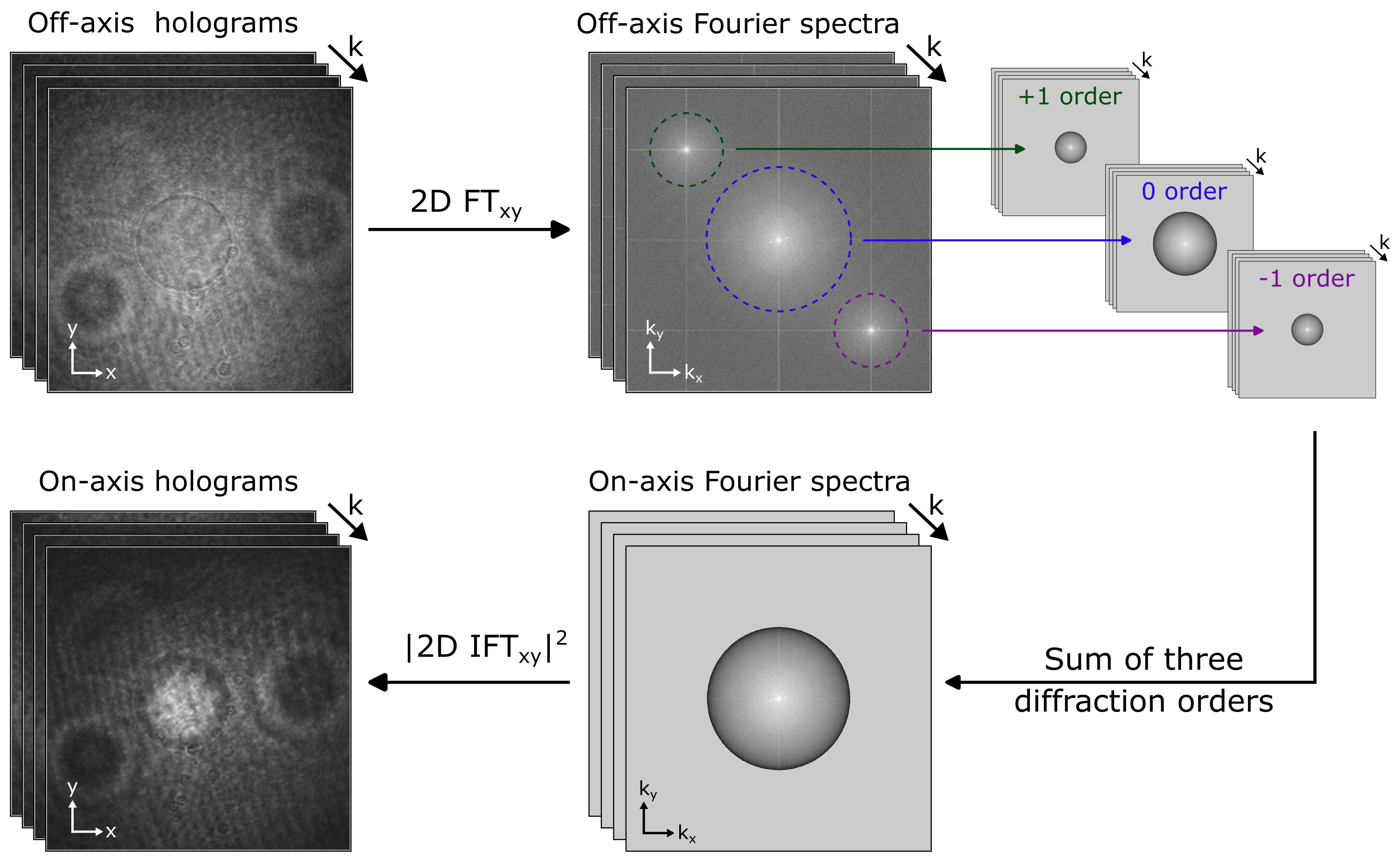}
\caption{Scheme for generating t-FF-SS-OCT holograms (spectral fringes) in on-axis configuration from t-FF-SS-OCT holograms captured in off-axis geometry. The t-FF-SS-OCT holograms were recorded for $751$ wavenumbers.}
\label{fig:scheme}
\end{figure}

The t-FF-SS-OCT (full-field swept-source OCT in transmission) system used in our work (see Method Section) operates in off-axis configuration, which allows for simultaneous retrieval of the phase with digital holography microscope (DHM) approach. The qtOCT approach, however, does not require spatial carrier frequency and can operate with on-axis geometry. To prove this, the off-axis t-FF-SS-OCT holograms were numerically transformed to on-axis t-FF-SS-OCT holograms. This procedure is depicted in Fig.~\ref{fig:scheme}. At first, 2D Fourier transform (FT) was calculated for each XY plane obtained for different wavenumbers $k$. In the resulting Fourier spectra 3 separated diffraction orders ($+1$, $0$, and $-1$ order) are visible, which is typical in the off-axis holography. Next, all diffraction orders have been isolated separately for each wavenumber using three main steps: 1) determination of the diffraction order diameter, 2) isolation of the diffraction order with circular Tukey window, 3) centering isolated region. Then, for each wavenumber, 3 isolated diffraction orders were summed, resulting in on-axis Fourier spectra. After that, the 2D inverse Fourier transform was calculated separately for each wavenumber followed by conversion of complex amplitudes to intensity images. Finally, all resulting images were stacked in a 3D matrix.

\end{document}